# Determination of flow stress and the critical strain for the onset of dynamic recrystallization using a sine function

Soheil Solhjoo[*]

Faculty of Mathematics and Natural Sciences, Advanced Production Engineering — Institute for Technology, Engineering & Management, University of Groningen, Groningen, The Netherlands

## Abstract

A new model has been developed to estimate the flow stress under hot deformation conditions up to the peak of the stress-strain curves. The model is constructed on the basis of the general form of sine functions by introducing an additional exponent. Besides, an equation is derived from the model, which could be used to predict the critical strain for the onset of dynamic recrystallization. The model has been examined for a 304 austenitic stainless steel. The estimated flow stress had an average percentage error of %1.66.

## 1. Introduction

In order to model the flow stress curves of metals and alloys during hot deformation processes, many different approaches are examined. However, all of them are functions of strain, strain rate and temperature [1]. Work hardening ($\theta = d\sigma/d\varepsilon$) and dynamic recovery control the flow stress up to a saturation stress. However, dynamic recrystallization (DRX) occurs before this maximum value which results in a single (or multi) peak stress in the flow curve. This mechanism also leads the flow stress to a steady-state value, lower than the peak stress. Although it is very important for designers to know the exact point of the onset of DRX, there are just a few mathematical attempts to determine this value [2-5] as a function of the deformation parameters, i.e. strain rate, and temperature.

---

*Email Address*: soheilsolhjoo@yahoo.com

Here, a new sine function is proposed to estimate the flow stress, which also could be used to predict the critical strain for the onset of DRX. This model is tested for '304 austenitic stainless steel' under different hot forming conditions. Calculated results are highly in accord with experiments.

## 2. Modeling of flow stress up to peak

Taking a general form of sine functions with an additional exponent ($K_1$), the following equation can be used for prediction of flow stress curves up to peak.

$$\sigma = \sigma_P \left[ \sin\left( \frac{\varepsilon}{\varepsilon_P} \frac{\pi}{2} \right) \right]^{K_1} \quad (1)$$

where $\sigma$ is stress, $\varepsilon$ strain, and the subscript P corresponds to the peak values. $K_1$, the adjustable process constant, is the slope of the linear plot of $\ln \sigma$ vs. $\ln \sin\left( \frac{\varepsilon}{\varepsilon_P} \frac{\pi}{2} \right)$, which can be described by:

$$K_1 = aT^{-b} \quad (2)$$

where T is the absolute temperature, and a and b are functions of strain rate. In order to find the values of a and b, natural logarithm of Eq.2 should be derived.

$$\ln K_1 = \ln a - b \ln T \quad (3-1)$$

$$\ln a = a_1 + a_2 \ln \dot{\varepsilon} \quad (3-2)$$

$$b = b_1 + b_2 \ln \dot{\varepsilon} \quad (3-3)$$

where $a_1$, $a_2$, $b_1$, and $b_2$ are constants, and $\dot{\varepsilon}$ is the strain rate.

## 3. Modification of the model

This model cannot be used in its original form for very low strains; $\varepsilon = 0$ results in $\sigma = 0$ instead of an initial stress value, $\sigma_0$. This limitation would be over by modifying Eq.1 by introducing a new parameter, $H_2$, as follows:

$$\sigma = \sigma_P \left[ \sin\left( \frac{\varepsilon + H_2}{\varepsilon_P} \frac{\pi}{2} \right) \right]^{K_1} \tag{4}$$

$H_2$ should change the predicted stress at $\varepsilon = 0$, from $\sigma = 0$ to $\sigma = \sigma_0$. However, it should not change the predicted value at the peak, because the original formula (Eq.1) would do that. In other words, $H_2$ should get defined in a way to satisfy these two points: (1) $\varepsilon = 0$, $\sigma = \sigma_0$ and (2) $\varepsilon = \varepsilon_P$, $\sigma = \sigma_P$. The following formula would satisfy these conditions (see Appendix A):

$$H_2 = (\varepsilon_P - \varepsilon) \frac{2}{\pi} \arcsin\left( \frac{\sigma_0}{\sigma_P} \right)^{\frac{1}{K_1}} \tag{5}$$

## 4. Critical strain for the onset of DRX

During hot deformation processes, the work hardening and dislocation density lead the material to a critical microstructural condition, where new grains nucleate and new high-angle boundaries grow. A driving force of dislocation removal initiates dynamic recrystallization, and let the flow stress increases with a declining rate, up to a maximum value (the peak), which indicates the rate of softening mechanism prevails over the work hardening. This phenomenon usually occurs for metals with low to medium stacking fault energy [6]. The influence of DRX would be more obvious as temperature rises or strain rate lowers [7].

The critical strain for the onset of dynamic recrystallization can be found metallographically. While it is time-consuming, the results are not precise [8]. Therefore, researchers put efforts to explain DRX mathematically. Perdrix and his colleagues [9] found that the $\theta - \sigma$ curves may be divided into successive linear portions with a negative slope up to the maximum point ($\sigma = \sigma_P, \theta = 0$). Ryan and McQueen [10] observed an inflection in the $\theta - \sigma$ curve corresponding to DRX initiation. Poliak and Jonas [8] showed that the minimum point of $\frac{d\theta}{d\sigma} - \sigma$ curve represents the onset of DRX.

Mathematically, the minimum point of $\frac{d\theta}{d\sigma} - \sigma$ is a null value of the second derivative of $\theta$ with respect to $\sigma$, and indicates the inflection of $\theta - \sigma$ curve [11].

$$\left. \frac{d^2\theta}{d\sigma^2} \right|_{critical\ strain} = 0 \tag{6}$$

Therefore, in order to solve Eq.4 the first step is to find the derivation of Eq.1[†] with respect to strain.

$$\theta = \frac{d\sigma}{d\varepsilon} = K_1 \frac{\pi}{2} \frac{\sigma_P}{\varepsilon_P} \cos\left(\frac{\varepsilon}{\varepsilon_P} \frac{\pi}{2}\right)\left[\sin\left(\frac{\varepsilon}{\varepsilon_P} \frac{\pi}{2}\right)\right]^{K_1-1} \quad (7)$$

By solving Eq.4 the critical strain for the onset of DRX ($\varepsilon_C$) could be found as a function of peak strain:

$$\varepsilon_C = \varepsilon_P \arctan\left(\sqrt{1-K_1}\right) \quad (8)$$

## 5. Results and Discussion

The experimental results of hot torsion tests of '304 austenitic stainless steel' from the literature [12] were used for the mathematical analyses in this investigation. As the very first step, the values of $K_1$ determined for all sets of deformation conditions. Using linear the plots of $\ln K_1$ vs $\ln T$, the values of $a$ and $b$ calculated for each set of strain rates. Fig. 1 illustrates the needed plots for calculation of $a$ and $b$ of Eq. 2, which in this study are determined to be as:

$$\ln a = 20.185 + 1.75 \ln \dot{\varepsilon} \quad (9-1)$$

$$b = 3.02 + 0.24 \ln \dot{\varepsilon} \quad (9-2)$$

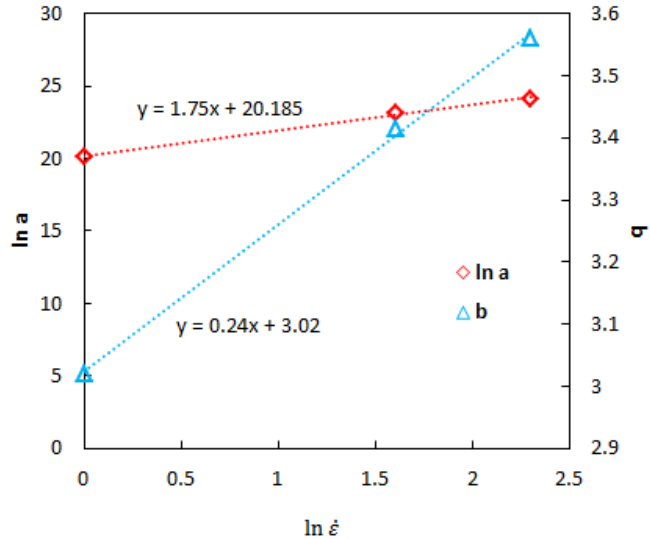

**Fig. 1.** *The relationship between ln a, b and* $\ln \dot{\varepsilon}$

---

[†] The derivation of the modified model (Eq.2) results in an indeterminable equation.

The calculated values of $K_1$ from both the experimental data and Eq. 2 are compared in Fig. 2, and shows a good correlation between them.

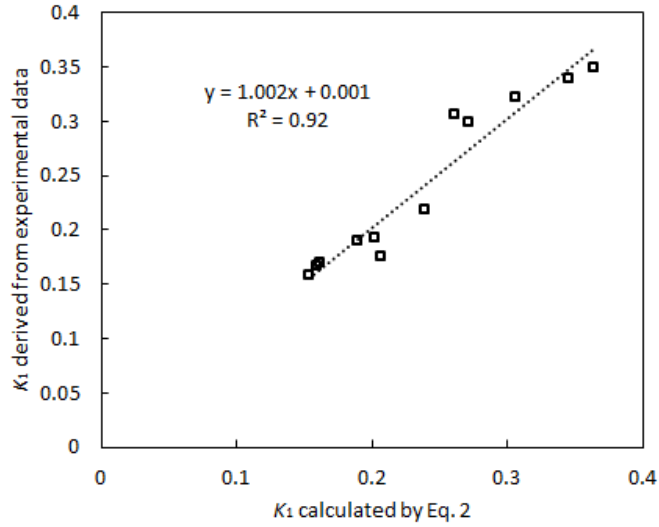

**Fig. 2.** *Comparison between the values of $K_1$ determined from experimental data and Eq.2*

Using Eq.4, flow stress is calculated for all different hot deformation conditions and compared with experimental results (for 370 data point). Fig.3 demonstrates this comparison.

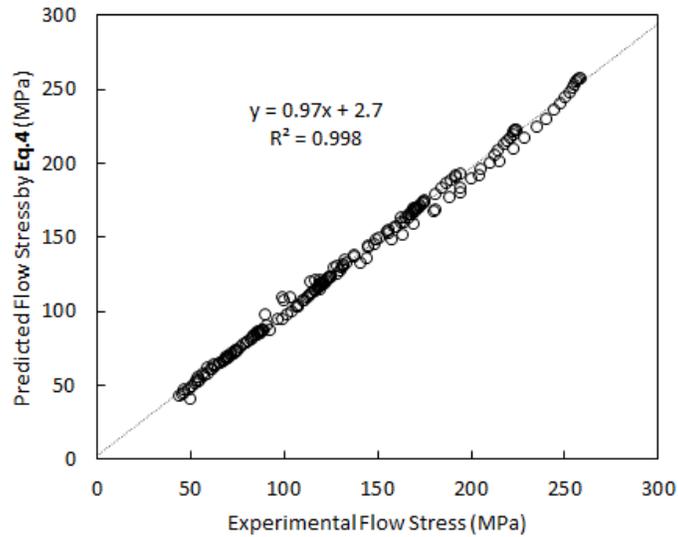

**Fig. 3.** *Comparison between experimental flow stress and predicted values by Eq.4*

The maximum, minimum and average percentage errors are evaluated as 15.33, 0.00 and 1.66, respectively. Afterwards, the coefficient of determination ($R^2$) is assessed by fitting a straight line to this diagram using the simple linear regression model. The formula and the $R^2$ value are reported in Fig.2. The slope of the fitted line is 0.97 with a coefficient of determination of 0.998. The low error values, the slope of the fitted line (≈1) and the high value of the coefficient of determination confirm that the calculated flow stress is highly in agreement with experiments.

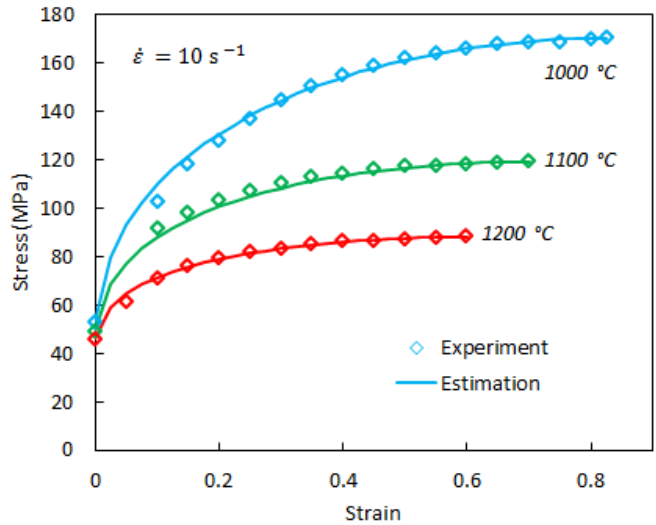

**Fig. 4.** *Comparison between experimental and predicted flow stress curves for a strain rate of 10 s$^{-1}$ at different temperatures*

Fig.4 depicts the flow stress curves for the strain rate of 10 s$^{-1}$, at different temperatures. The flow stress curves estimated by Eq.4, obviously, demonstrate a high accordance with experiments, right from the initial stress up to the peak. Fig.5 shows a plot of $\theta - \sigma$ for different strain rates at 900°C. In this figure, Eq. 5 is plotted for comparison with experimental data. The results show that the model is in agreement with experiments.

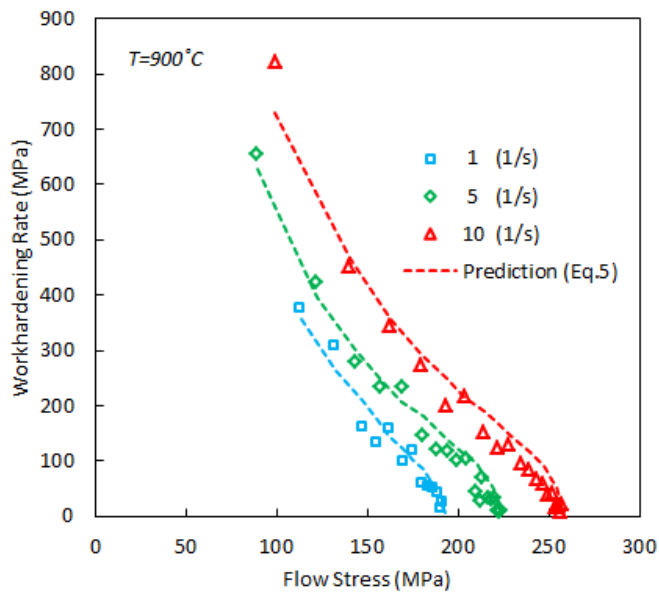

**Fig. 5.** *Comparison between experimental (points) and predicted (hashed lines) work hardening rate plots at 900°C with different strain rates*

Using Eq. 8 the values of $\varepsilon_c$ are calculated and compared with the experimental values [12] in Fig. 6.

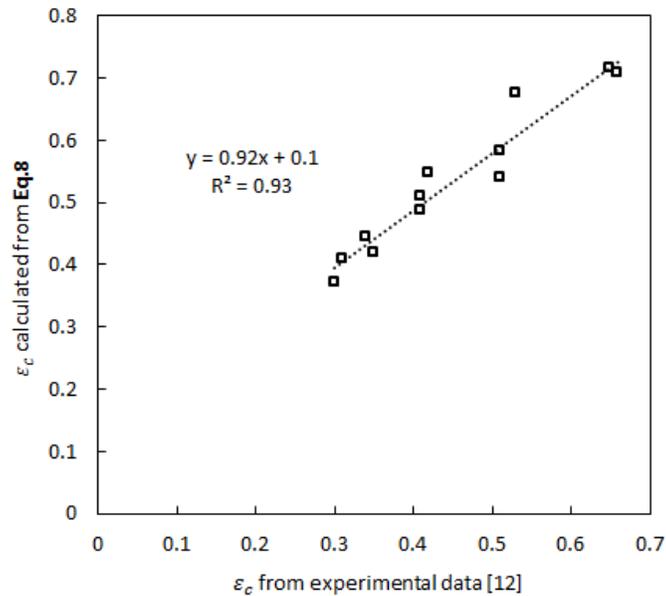

**Fig. 6.** *Comparison between experimental values of $\varepsilon_c$ [12] and calculated ones from Eq. 8*

## 6. Conclusion

A new constitutive model is developed to estimate the flow stress during the hot metal forming processes, up to the peak of the stress-strain curve. The model is based on the sine functions, with an additional adjustable exponent. The model is used to derive an equation which is able to predict the critical strain for the onset of dynamic recrystallization. The model tested for '304 austenitic stainless steel' and the results are highly accurate.

## Appendix A

In order to define $H_2$, first the values of $H_2$ at the boundary conditions are determined:

*Condition 1*: $\varepsilon = 0$, and $\sigma = \sigma_0$

$$\sigma_0 = \sigma_P \left[ \sin\left( \frac{0 + H_2}{\varepsilon_P} \frac{\pi}{2} \right) \right]^{K_1} \tag{1-A}$$

which results in:

$$H_2 = \varepsilon_P \frac{2}{\pi} \arcsin\left( \frac{\sigma_0}{\sigma_P} \right)^{\frac{1}{K_1}} \tag{2-A}$$

*Condition 2*: $\varepsilon = \varepsilon_P$, and $\sigma = \sigma_P$

$$\sigma_P = \sigma_P \left[ \sin\left( \frac{\varepsilon_P + H_2}{\varepsilon_P} \frac{\pi}{2} \right) \right]^{K_1} \tag{3-A}$$

which results in $H_2=0$.

In this work, I simply defined $H_2$ as a linear function of strain, which equals Eq.2-A at $\varepsilon = 0$, and zero at the peak strain. The following shows this definition:

$$H_2 = H_2(@ \varepsilon = 0) + \varepsilon \left[ \frac{H_2(@ \varepsilon = \varepsilon_P) - H_2(@ \varepsilon = 0)}{\varepsilon_P - 0} \right] \tag{4a-A}$$

$$H_2 = \varepsilon_P \frac{2}{\pi} \arcsin\left( \frac{\sigma_0}{\sigma_P} \right)^{\frac{1}{K_1}} + \varepsilon \left[ -\frac{2}{\pi} \arcsin\left( \frac{\sigma_0}{\sigma_P} \right)^{\frac{1}{K_1}} \right] \tag{4b-A}$$

Eq. 4b-A could be rewritten as:

$$H_2 = (\varepsilon_P - \varepsilon) \frac{2}{\pi} \arcsin\left( \frac{\sigma_0}{\sigma_P} \right)^{\frac{1}{K_1}} \tag{5-A}$$